\documentclass[prl,twocolumn,showpacs,preprintnumbers,amsmath,amssymb,floatfix]{revtex4}
\usepackage{epsfig}
\usepackage{graphicx}
\usepackage{amssymb}
\usepackage{color}
\usepackage{lipsum}

\begin{document}

\title{Noise-resistant optimal spin squeezing via quantum control}
\author{T.~Pichler$^{1}$*}
\author{T.~Caneva$^{1,3}$*}
\author{S.~Montangero$^1$}
\author{M.~D.~Lukin$^2$}
\author{T.~Calarco$^1$}

\affiliation{$^1$Institut f\"ur komplexe Quantensysteme, Universit\"at Ulm, D-89069 Ulm, Germany\\
$^2$Physics Department, Harvard University, Cambridge, Massachusetts 02138\\
$^3$ICFO-Institut de Ciencies Fotoniques, Av. Carl Friedrich Gauss, 3, 08860 Castelldefels, Barcelona, Spain}

\date{\today}
\begin{abstract}
Entangled atomic states, such as spin squeezed states, represent a promising resource for a new generation of quantum sensors and atomic clocks. We demonstrate that optimal control techniques can be used to 
substantially enhance the degree of spin squeezing in strongly interacting many-body systems, even in the presence of noise and imperfections. Specifically, we present a protocol that is
robust to noise which outperforms conventional methods.
Potential experimental implementations are discussed. 
\end{abstract}


\maketitle

Spin squeezed states are among the most interesting 
examples of entangled states. In quantum 
metrology they allow for measurements with an improved precision, 
ultimately limited only by the 
Heisenberg limit.
Since the early theoretical proposals to realize them with non linear interactions~\cite{Wineland_PRA92,Kitagawa_PRA93}, spin squeezed
states have been implemented in several experiments. Specific examples include generation of 
 spin squeezed states in cavity QED~\cite{Andre_PRL02,Sorensen_PRA02,Leroux_PRL10}, 
in trapped ions through shared motional 
modes~\cite{Molmer_PRL99,Leibfried_SCI04} 
or using a Bose-Einstein 
condensate~\cite{Gross_NAT10, Bucker_NP}.

\begin{figure}
\epsfig{file=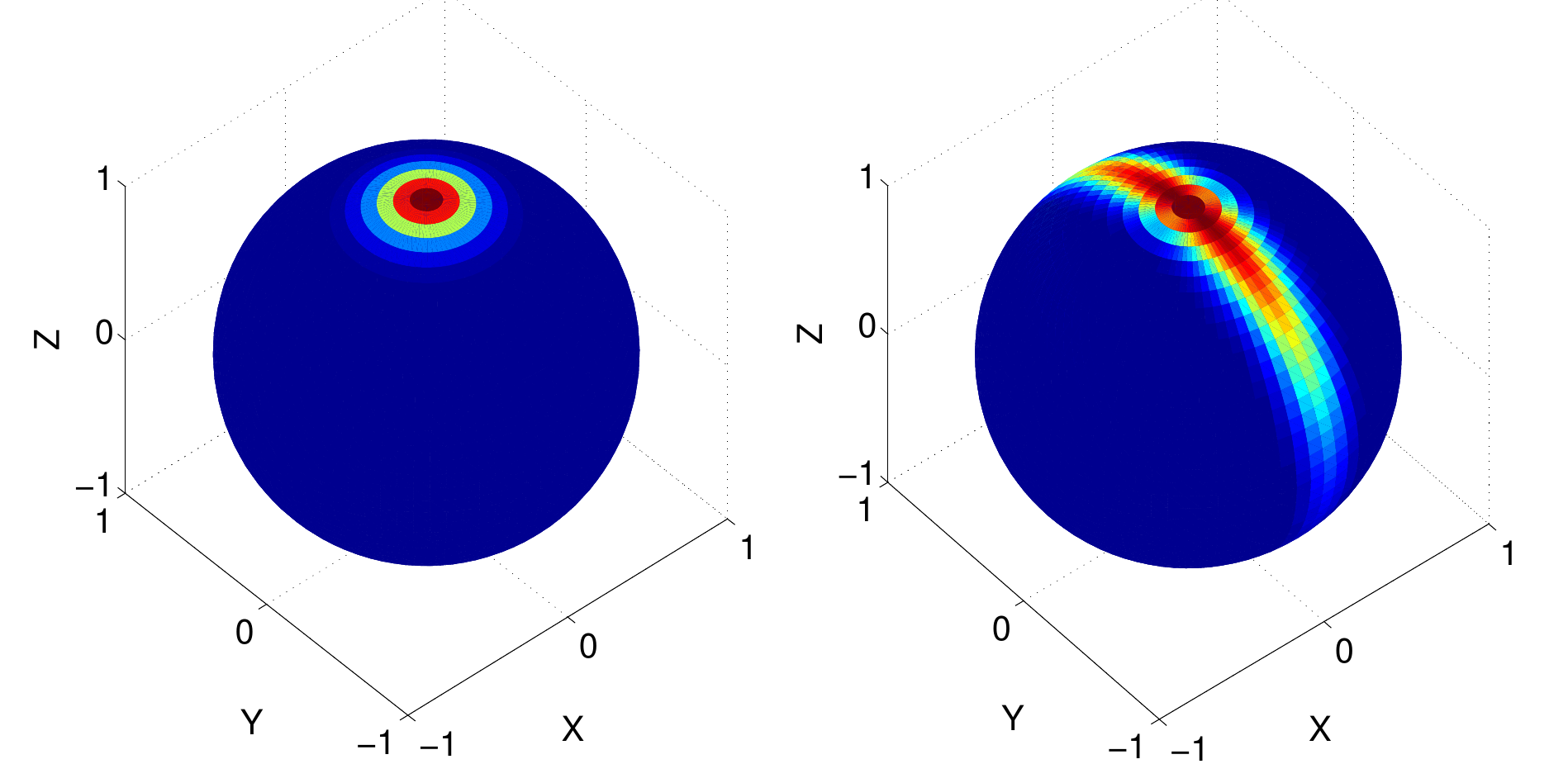,width=8cm,angle=0,clip=}
\epsfig{file=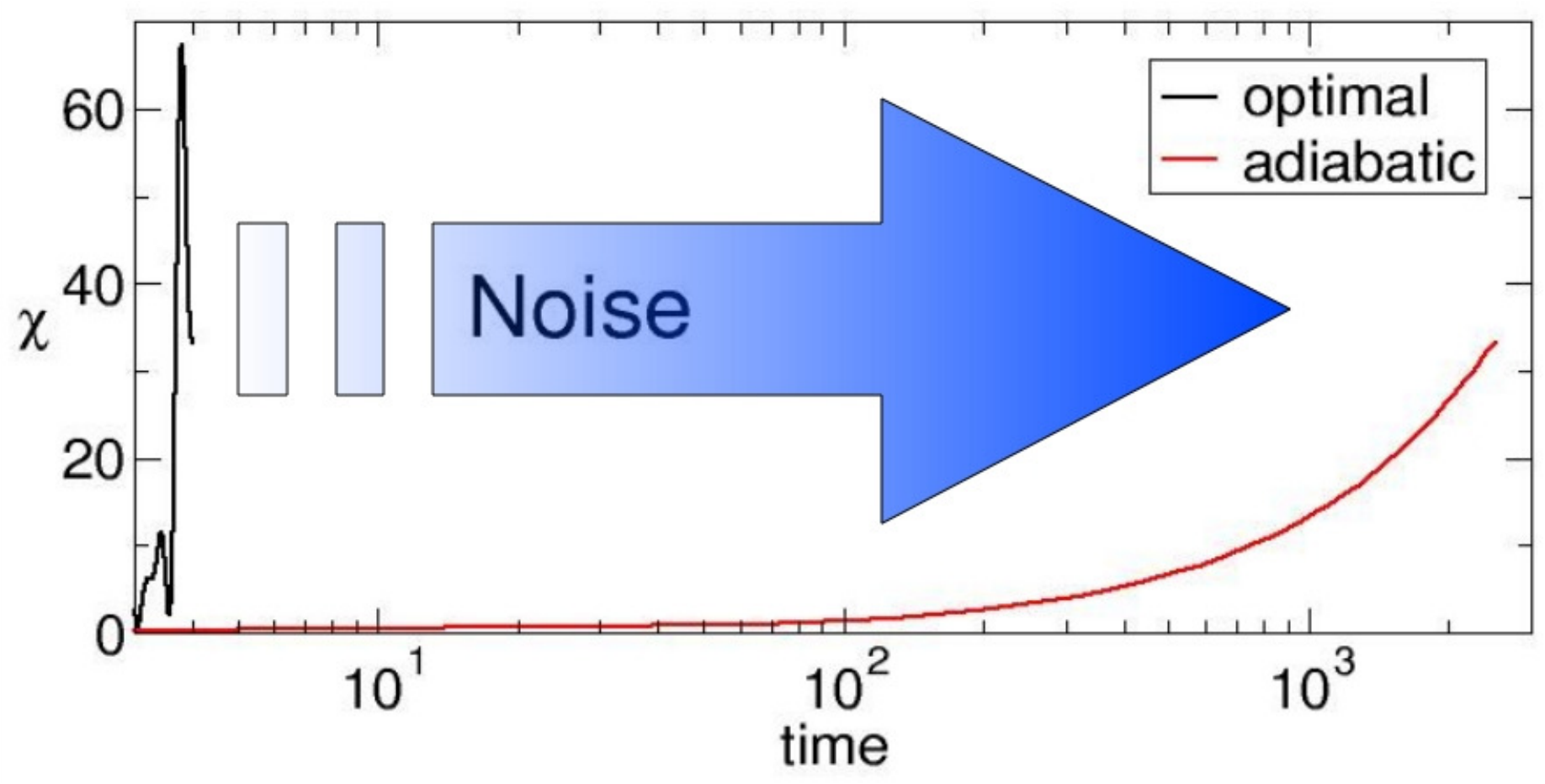,height=4cm,width=8cm,angle=0,clip=}
\caption{Upper panel: initial state (left) and 
final highly squeezed state (right) for a system of $N=100$
spins. Lower panel: adiabatic (red) and optimal (black) driving
fields $\chi$ generating the maximally squeezed state shown above; the effect of the 
noise (big blue arrow) increases with the total evolution time.}
\label{state_ini_fin:fig}
\end{figure}

In this Letter we demonstrate that optimal control can 
be effectively employed to produce highly squeezed spin states in
many-body quantum systems, drastically reducing the impact
of relaxation and decoherence. Other approaches applied control techniques creating spin squeezing as a succession of unitary pulses of a constant Hamiltonian \cite{Trail_PRL,Norris_PRL,Shen_PRA}.
 We employ the Chopped Random Basis (CRAB) technique \cite{CRAB,Caneva_PRA11} 
to optimally control the evolution of a 
collection of $N$ two-level systems mutually coupled through a time-dependent non 
linear (i.e. quadratic) interaction.
linear (i.e. quadratic) interaction. 
We calculate optimized evolutions occurring on time 
scales several orders of magnitude shorter than the corresponding 
adiabatic evolutions, with a speed-up increasing with the system size. Such a speed-up
translates directly into an enhanced robustness of the squeezing in the 
presence of noise, as schematically depicted in Fig.~\ref{state_ini_fin:fig}.
We illustrate this enhanced robustness by modelling two practical experimental 
implementations of squeezed state preparations: cavity QED and 
trapped ions~\cite{Leroux_PRL10,Leibfried_SCI04}. 

We will focus on two methods realizing spin squeezed states, both with 
advantages in different situations.
The first is based on the so called one-axis twisting protocol, 
consisting in letting a collection of two-level systems evolve
under the effect of a collective non linear interaction~\cite{Kitagawa_PRA93}, described by a Hamiltonian of the form
\begin{eqnarray}
 H_{SM}=\omega J_z+\chi J_x ^2
\label{ham_SM:eq}
\end{eqnarray}
Where $\omega$ is the precession frequency and $\chi$ is the strength of 
the nonlinear interaction and $J$ is a collective spin operator (defined below).
The relative simplicity of the one-axis twisting scheme has been at the 
basis of its ubiquitous presence in squeezing experiments; however
such a scheme is known to be non optimal~\cite{Kitagawa_PRA93},
the spherical nature of the angular momentum phase space limiting
the maximal squeezing achievable. Such a bound is intrinsic 
for the one-axis twisting protocol with fixed $\chi$. 
It nevertheless allows to achieve spin squeezing 
on comparably short time scales which makes it less sensitive with respect 
to noise.
The second protocol, proposed by S{\o}rensen and M{\o}lmer, 
is based on adiabatic evolution to 
steer a system into maximally squeezed states squeezing the variance 
$\Delta^2 J_x$ under the constraint
that $\langle J_z \rangle$ is nonzero.~\cite{Sorensen_PRL01}.

This procedure has been implemented experimentally in small
systems, see for instance Ref.~[\onlinecite{Chaudhury_PRL07}].
Unfortunately the required evolution time, which is proportional to the 
inverse square of the minimum 
spectral gap $\Delta$ encountered during the evolution, 
$T_{ad}\propto \Delta ^{-2}$~\cite{Messiah:book}, scales unfavourably with the system size. 
This makes adiabatic evolution significantly exposed to external noise: typically in many-body systems the gap closes 
with increasing system size $N$, which implies a dramatic increase of the time required for adiabatic
evolutions for large $N$.\\
Previous studies have demonstrated that optimal control 
is a powerful tool to drastically reduce the  
time needed to perform a  many-body quantum  
evolution~\cite{caneva_PRL09, Caneva_PRA11}.
In particular the Chopped Random Basis (CRAB) technique
offers an efficient way to implement optimal control, 
based on an expansion of the control field
onto a truncated basis~\cite{CRAB,Caneva_PRA11}. Recently it has been shown that  optimal control allows for reaching the Quantum Speed Limit (QSL), the minimal time required 
by physical constraints to perform a given transformation, in spin 
chains~\cite{caneva_PRL09,murphy_pra10}, cold atoms in optical lattices~\cite{bason_np11}, 
Bose-Einstein condensates in atom chip experiments
and in crossing of quantum phase transitions~\cite{caneva_pra11_1}.
Indeed, CRAB control makes it possible to reduce the time of the transformation 
down to the QSL, which scales as $1/\Delta$, obtaining a quadratic speedup of the 
protocol with respect to the adiabatic one. In this work we show that this method 
is successful also in drastically reducing the preparation time for maximally 
spin squeezed states, as illustrated in Fig.~\ref{state_ini_fin:fig}, 
thereby significantly enhancing the process' robustness to realistic 
noise sources even compared to the one-axis twisting protocol.

\emph{Model} ---
A collection of $N$ two-level atoms having (pseudo)spin $\vec{S}_i$ can be described in terms of the 
global spin variable
$\vec{J}=\sum _{i=1} ^N \vec{S}_i$,
with $|\vec{J}|=N/2$ and $z$-component $J_z$
representing the population imbalance between the two atomic internal 
states. 
In Ramsey spectroscopy experiments, the measured signal $M$ yields 
 the mean global angular momentum pointing along the $z$-axis, 
$M \equiv \langle J_z\rangle$, while the 
noise is given by the uncertainty in one of the orthogonal components
$\Delta J_i =\sqrt{\langle J_i ^2\rangle-\langle J_i\rangle ^2}$, $i=x,y$.
In spin squeezed states, the latter is below the standard quantum limit, i.e.
$\Delta{J_i}^2<|\langle J_j\rangle| /2$ for $i\neq j \in \{x,y,z\}$. 
The squeezing parameter $\xi$ is defined through 
the signal to noise ratio as
\begin{eqnarray}
 \xi =\frac{\sqrt{2J}\Delta J_x}{|\langle J_z\rangle|}.
\end{eqnarray}
Squeezed states satisfy the condition $\xi < 1$, which implies entanglement in the system.
The ideal states for spectroscopy experiments are those minimizing $\Delta J_x$ for sufficiently large values of the signal, 
i.e. $M\propto N$.
The problem of finding the optimal squeezed state can be recast
into the search for the ground state $|\psi_0(\chi,N)\rangle$ of the Hamiltonian Eq.~\ref{ham_SM:eq},
where $\omega$ is constant and negative and the non-linear interaction $\chi (t)$ is now taken to be tunable
in time ~\footnote{Indeed the problems
are equivalent for $J$ integer; instead for $J$ half-integer the 
procedure is allowed only if $\langle J_z\rangle/J$ 
is above a certain threshold, see \cite{Sorensen_PRL01}.}. (From now on we set 
$\hbar =1$ and  time is measured in units of $1/|\omega|$.)
Adiabatic evolution under $H_{SM}$ automatically produces optimal squeezed states, as follows:
At the time $t=0$ one takes $\chi (0)=0$
and the system is prepared in its initial ground state $|\psi_0(0,N)\rangle$, the coherent state 
$|J_z =J\rangle$ with $\xi =1$.
Then adiabatically increasing $\chi (t)$, the system evolves
following the instantaneous ground state $|\psi_0(\chi(t),N)\rangle$ of $H_{SM}$, yielding exactly  the 
family of states with optimal squeezing at a given 
value of $M$ 
(see Fig.~\ref{state_ini_fin:fig}). 
%
\begin{figure}
\epsfig{file=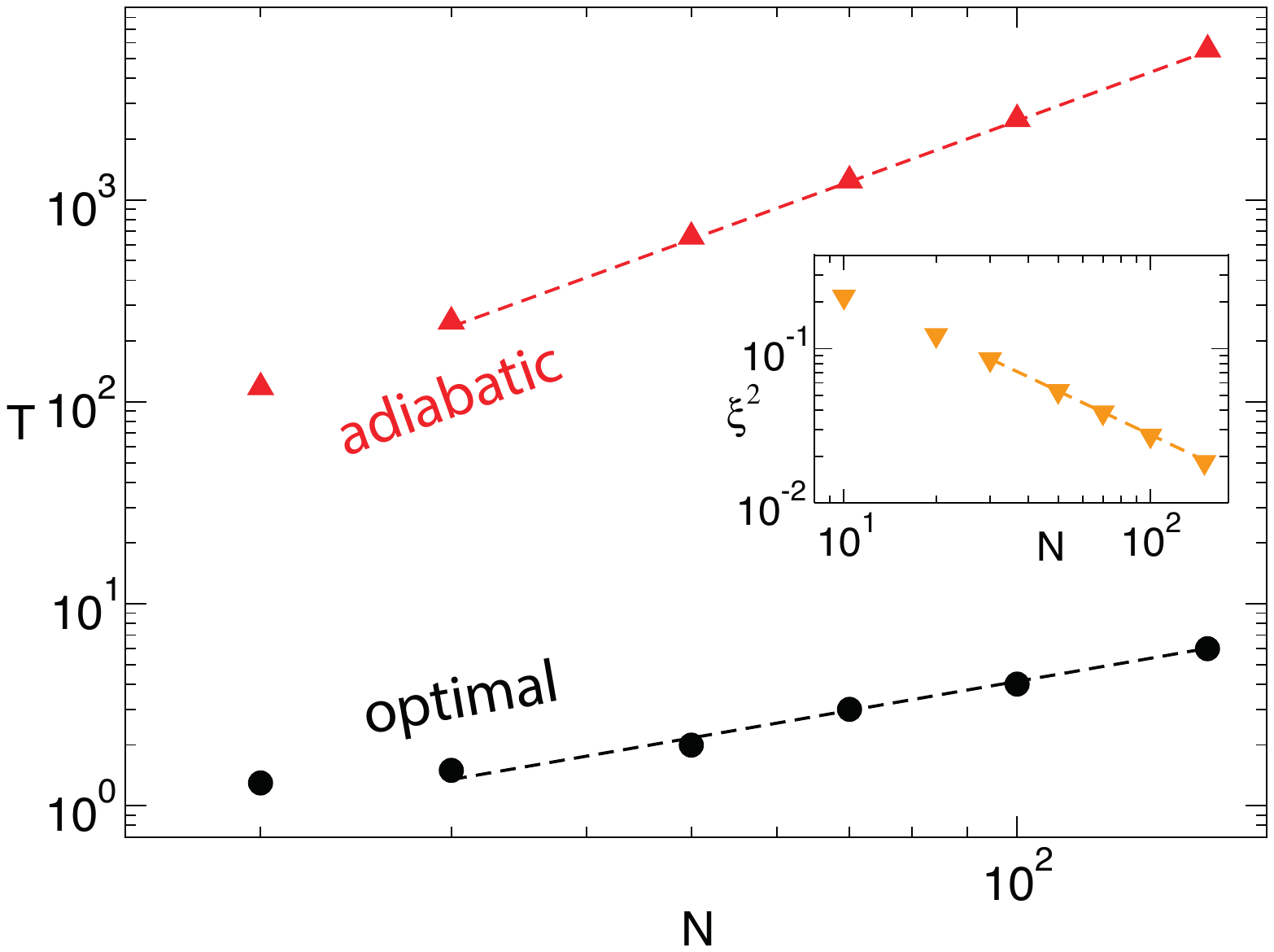,width=7cm,angle=0,clip=}
\caption{Scaling with the size of the total 
evolution time $T$ for the adiabatic ($I_{ad} = 7 \cdot 10^{-3}$, red triangles) and the optimized dynamics 
($I_{opt} = 5 \cdot 10^{-4}$, black circles). Numerical fits
for $30\leq N\leq 150$ (dashed lines) result in $T_{ad}\propto N^{1.95}$ and $T_{opt}\propto N^{0.93}$. Inset: Scaling of $\xi ^2$; a fit gives
$\xi ^2\sim 2.1/N^{0.94}$.}
\label{statics_dynamics:fig}
\end{figure}

\emph{Optimization in the absence of noise} ---
We first investigate the properties of the Hamiltonian $H_{SM}$
in Eq.(\ref{ham_SM:eq}) to identify target squeezed states that can be reached via adiabatic evolution. We calculate the time required to perform an adiabatic transformation from the initial 
state into the target and its scaling with the system size $N$. Subsequently, 
we apply optimal control to determine the dynamics  
(neglecting for the moment decoherence effects) 
leading to the same target state in a much shorter time. Finally, we compare the optimized evolution with 
the adiabatic one. 

As previously mentioned, squeezed states suitable for quantum metrology 
should have sufficiently strong signal $M$. 
To fulfill this requirement we choose (throughout the whole work)  
$\bar M=J/\sqrt{2}=0.707J$, i.e. $ \bar M \propto N$. 
Then we find the value $\chi_{\bar M}(N)$ of the interaction such that 
$|\psi_0(\chi_{\bar M},N)\rangle$ has $\langle J_z \rangle = \bar M $ for a given $N$.
The inset of Fig.~(\ref{statics_dynamics:fig}) shows the corresponding value of the ground-state squeezing 
for varying $N$: 
a power-law fit $\xi ^2=A/N^B$ for $30\leq N\leq 150$ yields $A=2.1 \pm 0.05$ 
and $B=0.94 \pm 0.01$, compatible with the Heisenberg limit $\xi ^2\propto N^{-1}$.
This means that we have identified a class of states $| \psi_0(\chi_{\bar M},N) \rangle$ with the desired characteristics. We can now take those states as a target for the optimization, to achieve 
constant intensity of the signal $\bar M$ and maximal squeezing $\xi$ for any given system size $N$.

As discussed above, the system is initially prepared 
in the coherent state $| \psi_0(0,N) \rangle$ 
where all spins are polarized along the positive $z$-direction and $\xi ^2=1$, and we aim at reaching the goal state $|\psi _G\rangle \equiv | \psi_0(\chi_{\bar M},N) \rangle$ after an evolution time $T$. 
The initial and target state for the case $N=100$ are depicted in Fig.~\ref{state_ini_fin:fig} 
(upper panels).
For the adiabatic case, evolution is computed  using a linear ramp $\chi(t) =\chi_{\bar M}  t/T$. Comparing the resulting final state $| \psi(T) \rangle$ with 
the goal state yields the infidelity $I =1-|\langle \psi_0(\chi_{\bar M},N) | \psi(T) \rangle |^2$.
Fig.~\ref{statics_dynamics:fig} shows, as a function of the size $N$, the time $T_{ad}$ needed to reach a given infidelity value $I_{ad}$  via adiabatic evolution (red triangles). 
A fit $T=AN^B$ for $30\leq N\leq 150$ gives $A=0.31 \pm 0.01$ and $B=1.95 \pm 0.01$, in agreement with the 
prediction of the adiabatic theorem $T_{ad} \sim 1/\Delta^2 \sim 1/N^2$.
We then apply the quantum optimal control CRAB algorithm~\cite{CRAB,Caneva_PRA11} to find the time $T_{opt}$ needed by an optimal transformation  to reach
an infidelity $I_{opt}$.
More precisely we write the driving field in the form 
$\chi(t) = \chi_{\bar M}   [1 + \lambda(t) \sum_{j=1}^{n_f} a_j \sin( \omega_j t) + b_j \cos( \omega_j t)]  t/T$, 
where $\lambda(t)$ ensures constant boundary conditions, 
$\omega_j = 2 \pi/T (1+r_j)$, $r_j$ is a random number, and $n_f \sim O(10)$, and we look for the optimal correction (i.e. the coefficients $\vec a, \vec b$) such that the infidelity is minimised for a given time (for details on the algorithm and 
of its complexity see~\cite{Caneva_PRA11,caneva_inprep}). A typical result is shown in Fig.~\ref{state_ini_fin:fig} 
(lower panel), while the scaling of the optimized evolution time $T_{opt}$ as a function of the size $N$ is shown 
in Fig.~\ref{statics_dynamics:fig} (black circles). A power-law fit 
$T_{opt}=AN^B$ for $30\leq N\leq 150$ gives $A=0.06\pm 0.01$ and $B=0.93 \pm 0.04$, consistent with our conjecture about the QSL (see above). This shows that optimal squeezing preparation results in 
a quadratic improvement in the scaling of the preparation time as a function of the system 
size, while additionally reducing the total evolution time by at least two orders 
of magnitude.
\begin{figure}
\epsfig{file=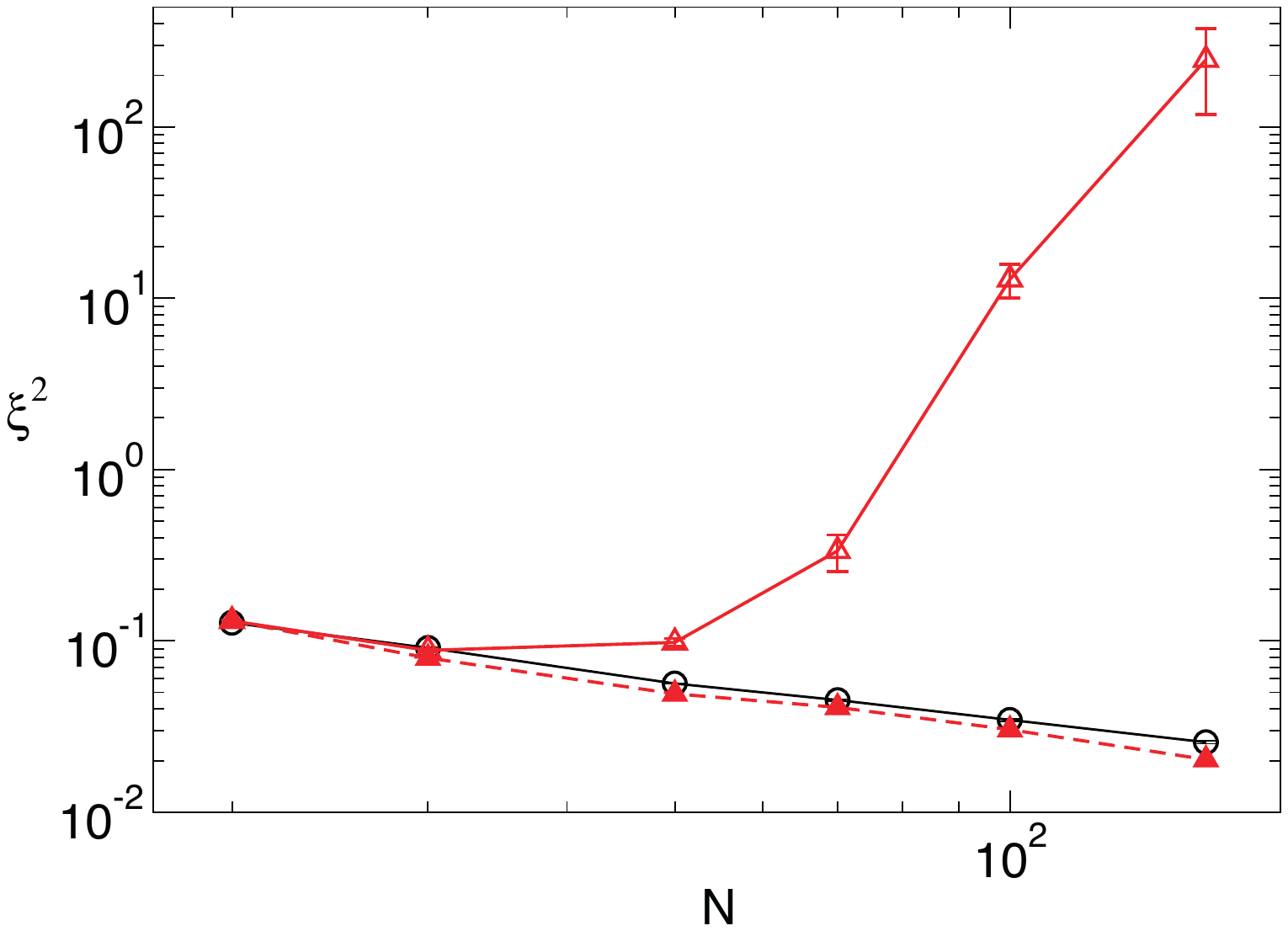,width=8cm,angle=0,clip=}
\caption{Final squeezing $\xi ^2$ as a function of the size $N$, for $\nu=500$ 
for the adiabatic (red triangles) and optimal (black circles) dynamics, subject to random telegraph noise with amplitude $K_{\alpha}=K_{\beta}=0.05$ 
(empty symbols) and $K_{\alpha}=K_{\beta}=0$ (full symbols). 
Data have been averaged over 24 instances of disorder.
}
\label{noise:fig}
\end{figure}

Our discussion up to this point neglected completely the effect of noise, which of course is a major concern in a real experiment. Therefore, in order to test the robustness of the protocol, we simulate the dynamics of the system in the presence of noise. We will consider two noise models, as different experimental implementations of squeezed spin states are affected by different kinds of noise. We will show that  optimized protocols work also in the presence of these types of noise, and that they are much more resilient to noise than adiabatic protocols.

\emph{Effect of classical noise} ---
A typical situation in which classical fluctuations of an external field occur, reflecting in random fluctuations of the interaction strength, is found in trapped ions, also relevant for metrological applications~\cite{Leibfried_SCI04}. In trapped ion systems, a global random magnetic 
noise is expected to be the most relevant source of disturbance~\cite{Monz_PRL11}. 
We  include it in our simulations by adding random classical telegraph noise 
to the control field. We then study the evolution induced by the Hamiltonian 
\begin{eqnarray}
H=\chi (t)[1+K_{\alpha}\alpha (t)]J_x^2+ \omega[1+K_{\beta}\beta (t)]  J_z
\label{ham_noise:eq}
\end{eqnarray}
where $\alpha (t),\beta (t)$ are random functions of the time with a flat 
distribution in $[-1,1]$, changing random value on average with frequency $\nu$.
The case $K_\alpha= K_\beta=0$ corresponds to a noiseless evolution of 
Eq.\eqref{ham_SM:eq}. In Fig.~\ref{noise:fig} we compare the 
effect of the noise on the final squeezing obtained by varying $\chi (t)$ either linearly in time
(empty red triangles) or according to the optimized protocol 
(empty black circles).
The squeezing $\xi ^2$ is plotted as a function of the size $N$, for $\nu=500$ 
and for an intensity of the noise $K_{\alpha}=K_{\beta}=0.05$.
As shown in Fig.~\ref{noise:fig}, the noise effect is stronger for larger system sizes, 
very quickly destroying the squeezing for the slow linear (adiabatic) protocol. 
The reason is simple: as shown in Fig.~\ref{statics_dynamics:fig}, 
for large sizes, e.g. $N\geq 100$, the adiabatic evolution time is three orders
of magnitude larger than the optimized one. Vice versa 
the fast optimal driving turns out to be robust even at large sizes and relatively 
high intensities of the noise, resulting in a final squeezing almost 
equivalent to that obtained via the adiabatic process in the absence of noise (full red triangles). 

\emph{Effect of quantum noise} ---
Recently, techniques in evolving interactions of spin ensenbles with nano-mechanical resonators have investigated the possible implementation of the one-axis twisting protocol, showing comparable results to ours in a similar range of the collective cooperativity \cite{ben13}.\\
Finally we discuss a noise model suitable for the description of QED experiments \cite{Leroux_PRL10},
in which the effect of the noise is treated through the formalism
of the master equation. In cavity QED, relaxation of the atomic levels towards the ground 
state and  leakage of photons outside the cavity are  the most
relevant source of dissipation~\cite{Sorensen_PRA02}. 
In order to  estimate the effect of the noise in a realistic system, we derive the Hamiltonian of Eq.~(\ref{ham_SM:eq})
from a microscopical model.  
\begin{figure}[t]
\epsfig{trim=2cm 7.5cm 2cm 8cm, file=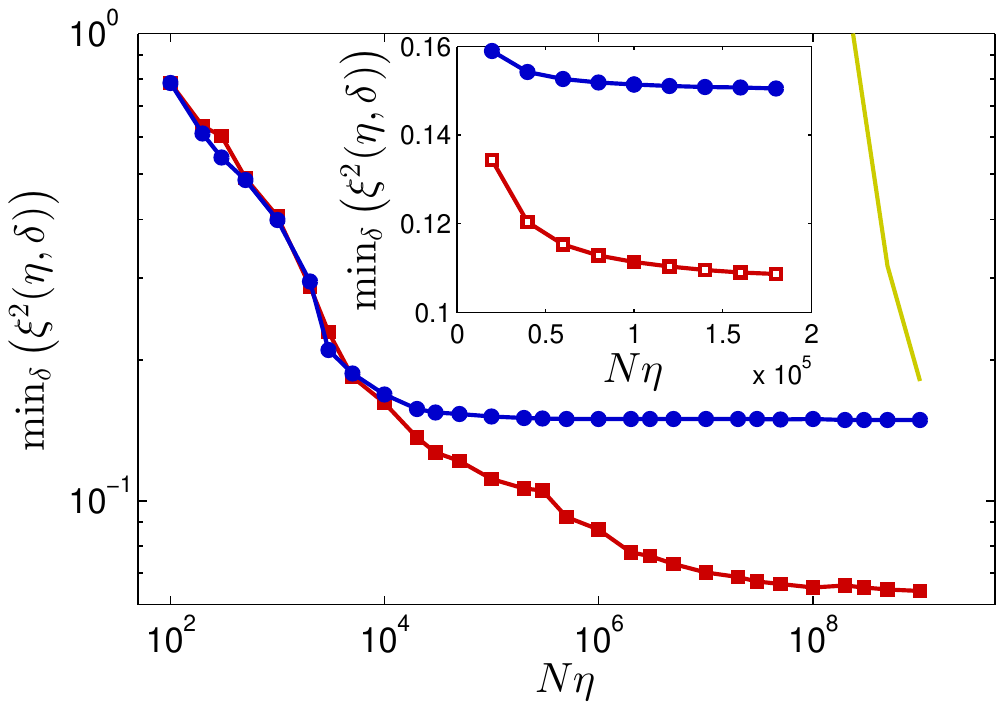,width=8cm,angle=0,clip=}
\caption{
Squeezing $\xi^2$
as a function of the collective cooperativity $N\eta$ in the case of one-axis 
twisting (blue circles), optimal (red squares) and adiabatic 
(yellow) protocols with $N=30$.\\
Inset: $\xi^2$ as a function of $N\eta$
for the one-axis twisting (blue) compared to the optimized pulse 
obtained at $N\eta=10^5$ (full square) and applied for different values of the cooperativity (empty squares).
}
\label{coop_maseq_sqx_vs_int_N30:fig}
\end{figure}
We consider a collection of $N$ three level atoms with two stable 
ground states $|a\rangle$ and $|b\rangle$ and an excited state
$|e\rangle$, in an optical cavity; the ground state energy splitting
is given by $\omega _{ab}$ and the relevant cavity mode has
a frequency $\omega _0$.   
The stable ground state $|a\rangle$ ($|b\rangle$) is coupled to the excited state
with a Rabi frequency $\Omega _1$ ($\Omega _2$) and a frequency $\omega _1$
($\omega _2$) which is detuned from the excited state by $\Delta _1$
($\Delta _2$). 
In the regime of weak laser power, the excited level is almost not populated 
and it can be adiabatically eliminated, leading to an effective 
photon-mediated interaction between the two ground state levels $|a\rangle$ and
$|b\rangle$. 
By introducing the total angular momentum operators 
$J_+=\sum _{k=1}^N |a\rangle _k  \langle b| _k$, 
$J_-=\sum _{k=1}^N |b\rangle _k  \langle a| _k$
and $J_z=(\sum _{k=1}^N |a\rangle _k  \langle a| _k -|b\rangle _k  \langle b| _k)/2$,
and by further assuming the strength of the two Raman processes 
to be identical,
$\Omega _1 g_b^*/\Delta _1=\Omega _2 g_a^*/\Delta _2=\Omega g^*/\Delta$,
after adiabatically eliminating also the cavity field,
we obtain the following  
master equation for the density matrix~\footnote{see the Supplementary Material for details}:
\begin{eqnarray}
 \dot{\rho}=-i[\tilde{H}_{eff},\rho ]+\mathcal{L}\rho ,
\label{master_eq:eq}
\end{eqnarray}
with unitary part given by
\begin{eqnarray}
\tilde{H} _{eff}= \omega J_z + \chi J_x ^2 ,
\end{eqnarray}
where $\chi=|\Omega|^2|g |^2/\delta\Delta ^2$ and 
$\delta=\omega _1-\omega _0 -\omega _{ab}$,
and nonunitary part described by the Linbladian
\begin{eqnarray}
\label{L:eq}
 \mathcal{L}\rho &=& \tilde{\gamma} [2J ^{\dagger}\rho J ^{-} -J^-J ^{\dagger}\rho-\rho J^-J ^{\dagger}],
\end{eqnarray}
where, from a microscopical derivation of the model, the most relevant contribution to the relaxation rate is $\tilde{\gamma}= \chi (t)\gamma\delta/|g^2|$. 

%

\emph{Experimental implications} ---
As discussed above, the Hamiltonian Eq.~(\ref{ham_SM:eq}) is relevant e.g. for the experiment 
of Ref.~\cite{Leroux_PRL10}. Here, squeezing of the collective spin of atoms in a cavity is used to improve the measurement precision of an atomic clock. 
With a realistic estimate of the parameters \cite{Leroux_PRL10} we have
$\Delta = 780$ nm $\sim 3\cdot 10^{14}$~Hz,
$\delta \sim  2\pi\cdot 3$~GHz,
$\gamma \sim  2\pi \cdot 5$~MHz,
$g \sim  2\pi \cdot 0.4$~MHz,
$\kappa \sim  2\pi \cdot 1$~MHz.
The dominant part for the relaxation
is thus proportional to the intensity of driving field $\chi $
with a proportionality constant given by $ \gamma\delta/|g|^2\sim 10^5$. 
Our estimate of relaxation rate can be also expressed in terms of cooperativity $\eta =g^2/(\gamma\kappa)$, leading to
$\tilde{\gamma}=\chi (t)\delta/(2\kappa\eta)$, where 
$\kappa$ is the decay rate of the cavity.
In Fig.~\ref{coop_maseq_sqx_vs_int_N30:fig} 
the squeezing parameter is shown as a
function of the collective cooperativity $N\eta$, for a system of size $N=30$. 
We compare optimized results directly obtained for different values 
of the cooperativity with values achieved with the 
one-axis twisting protocol which is known to be robust with respect 
to noise. 
We found that for values higher than $N\eta=10^4$ we can achieve better 
results for the squeezing parameter improving further as the value of the 
cooperativity gets increased, a behavior we observe also in simulations with different $N$.\\
Additionally, we compared the results of the optimized pulses with the 
adiabatic protocol of S{\o}rensen and M{\o}lmer \cite{Sorensen_PRL01} which
achieves optimal squeezing for long time-scales and high cooperativities.
In Fig.~\ref{coop_maseq_sqx_vs_int_N30:fig} some comparative results of the 
adiabatic protocol are shown. Further results of our simulations have shown that 
with optimized pulses the same results are achievable as with the adiabatic protocol 
at a cooperativity seven orders of magnitude higher. This is a large improvement towards 
optimal squeezing at practical accessible values of the cooperativity.\\
The inset of Fig.~\ref{coop_maseq_sqx_vs_int_N30:fig} displays the stability of 
of a certain optimal pulse in the high-noise regime. Here we used the pulse 
obtained for a cooperativity of $N\eta=10^5$ and applied it for a wide range of
different values of $N\eta$. Throughout these values the same optimal pulse 
improves the squeezing in comparison to the one-axis-twisting protocol.

\emph{Conclusions and outlook} ---
We 
have shown that optimal control can be used to speed up the dynamics for 
the production of squeezing with an additional improvement in the scaling 
of the preparation time as a function of the system size.
Also we have demonstrated that optimized evolutions scale better 
with noise than the one-axis twisting protocol providing the best 
values of squeezing known in this context.
In fact, in comparison with the adiabatic protocol, 
we were able to achieve maximally squeezed 
states a lot more robust 
with respect to the noise than with the adiabatic protocol.
The implementation of optimized protocols 
in spin squeezing experiments could therefore have a great impact in the field of 
quantum metrology. 
The implementation of closed-loop optimal control 
strategies might result in additional improvement~\cite{inguscio_inprep}. 
Finally, application of the present methods demonstrated here to more complex 
spin squeezing schemes~\cite{Andre_PRL02}, as well as adiabatic quantum 
computation in the presence of decoherence, can also be envisioned.

We thank F.~Schmidt-Kaler for discussions 
and 
NSF, CUA, ARO MURI, Packard Foundation, the SFB TRR21 and the European Commission through grants QIBEC 
and SIQS for support.\\
* These two authors contributed equally to this paper.




\begin{thebibliography}{99}

\bibitem{Chuang:book} M. Nielsen and I. L. Chuang, \emph{Quantum Computation and Quantum Information}, Cambridge University Press (2000) 

\bibitem{Wineland_PRA92} D.~J.~Wineland, J.~J.~Bollinger, W.~M.~Itano, F.~L.~Moore, and D.~J.~Heinzen Phys. Rev. A {\bf 46}, 6797 (1992)

\bibitem{Kitagawa_PRA93} M. Kitagawa and M. Ueda, Phys. Rev. A {\bf 47}, 5138 (1993)

\bibitem{Sorensen_PRA02} A.~S. S{\o}rensen and K. M{\o}lmer, Phys. Rev. A {\bf 66}, 022314 (2002)

\bibitem{Andre_PRL02} A. Andr\'e, L.~M. Duan, and M.~D. Lukin, Phys. Rev. Lett. {\bf 88}, 243602 (2002)

\bibitem{Leroux_PRL10}  I.~D. Leroux, M.~H. Schleier-Smith, and V. Vuletic, Phys. Rev. Lett. {\bf 104}, 073602 (2010)

\bibitem{Molmer_PRL99} K. M{\o}lmer and A.~S. S{\o}rensen, Phys. Rev. Lett. {\bf 82}, 1835 (1999)

\bibitem{Leibfried_SCI04} D. Leibfried et al, Science {\bf 304}, 1476 (2004)

\bibitem{Gross_NAT10} C. Gross, T. Zibold, E. Nicklas, J. Est\'eve, and M. Oberthaler, Nature {\bf 464}, 1165 (2010)

\bibitem{Bucker_NP} R. B\"ucker, J. Grond, S. Manz,	
T. Berrada, T. Betz, C. Koller, U. Hohenester, T. Schumm,	
A. Perrin, and J. Schmiedmayer, Nature Physics {\bf 7}, 608 (2011)


\bibitem{Trail_PRL} C. M. Trail, P. S. Jessen, and I. H. Deutsch, Phys. Rev. Lett. {\bf 105}, 193602 (2010)

\bibitem{Norris_PRL} L. M. Norris, C. M. Trail, P. S. Jessen, and I. H. Deutsch, Phys. Rev. Lett. {\bf 109}, 173603 (2012)

\bibitem{Shen_PRA} C. Shen, and L.-M. Duan, Phys. Rev. A {\bf 87}, 051801(R) (2013)



\bibitem{CRAB} P. Doria, T. Calarco, and S. Montangero, Phys. Rev. Lett. {\bf 106}, 190501 (2011)

\bibitem{Caneva_PRA11} T. Caneva, T. Calarco, and S. Montangero, Phys. Rev. A {\bf 84}, 022326 (2011)

\bibitem{Sorensen_PRL01} A.~S. S{\o}rensen and K. M{\o}lmer, Phys. Rev. Lett. {\bf 86}, 4431 (2001) 

\bibitem{Chaudhury_PRL07} S. Chaudhury \emph{et al.}, Phys. Rev. Lett. {\bf 99}, 163002 (2007)

\bibitem{Messiah:book} A. Messiah, \emph{Quantum Mechanics}, North-Holland (1962) 

\bibitem{caneva_PRL09}  T. Caneva, M. Murphy, T. Calarco, R. Fazio, S. Montangero, V. Giovannetti, and G. E. Santoro, Phys. Rev. Lett. {\bf 103}, 240501 (2009)

\bibitem{murphy_pra10} M. Murphy, S. Montangero, V. Giovannetti, and T. Calarco, Phys. Rev. A {\bf 82}, 022318 (2010)

\bibitem{bason_np11} M. G. Bason, M. Viteau, N. Malossi, P. Huillery, E. Arimondo, D. Ciampini, R. Fazio, V. Giovannetti, R. Mannella, and O. Morsch, Nature Physics {\bf 8}, 147 (2011)

\bibitem{Monz_PRL11} T. Monz et al, Phys. Rev. Lett. {\bf 106}, 130506 (2011)

\bibitem{caneva_pra11_1} T. Caneva, T. Calarco, R. Fazio, G. E. Santoro, and S. Montangero, Physical Review A {\bf 84}, 012312 (2011)

\bibitem{caneva_inprep} T. Caneva, A. Silva, T. Calarco, R. Fazio, S. Montangero, Phys. Rev. A {\bf 89}, 042322 (2014)

\bibitem{Lipkin_NP65} H.~J. Lipkin, N. Meshkov, and A.~J. Glick, Nucl. Phys. {\bf 62}, 188 (1965) 

\bibitem{Botet_PRB83} R. Botet and R. Jullien, Phys. Rev. B {\bf 28}, 3955 (1983)

\bibitem{Larson_EPL10} J.~Larson, EPL {\bf 90}, 54001 (2010)

\bibitem{Smith_PRA10} M.~H. Schleier-Smith, I.~D. Leroux, and V. Vuletic, Phys. Rev. A {\bf 81}, 021804(R) (2010)

\bibitem{ben13}S.D. Bennet, N.Y. Yao, J. Otterbach, P. Zoller, P. Rabl, and M.D. Lukin, Phys. Rev. Lett. {\bf 110}, 156402 (2013)

\bibitem{inguscio_inprep} S. Rosi, A. Bernard, N. Fabbri, L. Fallani, C. Fort, M. Inguscio, T. Calarco, S. Montangero, Phys. Rev. A {\bf 88}, 021601(R) (2013)





\end{thebibliography}
\end{document}